\documentclass[aps,preprint]{revtex4}
\begin{document}
\begin{flushleft}
Proceedings I Einstein Centenary Symposium, 303-307 
\end{flushleft}
\title{On The Origin of Universe}
\author{S.C. Tiwari}
\affiliation{Central Electronics Engineering Research Institute, \\
Pilani-333 031 (Rajasthan) India}
\maketitle

Amongst various theories concerning the large scale structure of 
the Universe and its origin, the big-bang or evolutionary theory 
and steady state theory have been extensively discussed and 
developed. According to big-bang idea, the Universe started from 
a very hot and dense (infinite density) matter with the greatest 
expansion rate at the beginning and then slowing down steadily. 
On the other hand, steady state theory assumes that the expansion 
has always been going on at just its present rate and the matter 
is continually created to compensate for the background material. 
These theories are, however, inadequate to answer the following 
questions: What was the state before the creation of the 
Universe ? And, where does the Universe expand ? In the first 
model, the question that what happened before the big-bang can 
not be answered. In steady state theory the question cannot be 
asked meaningfully since it is assumed that the Universe as 
observed from a particular position and at a particular epoch 
is same as that observed from any other position and at any other 
epoch. The matter is postulated to be created from nothing. 
The status of the answer to the second question is very much 
unclear and confused in both theories.

An attempt is made to answer some of these questions starting from 
a new hypothesis \cite{1} on space-time interaction contained in 
two postulates stated below: 
\begin{itemize}

\item[(a)] There is a state, 'Superspace'  which has its own energy 
called fundamental energy, 

\item[(b)] Time is something which operates upon the state 
'Superspace' to transform the fundamental energy to known form of 
energy thereby creating Universe. 
\end{itemize}

It would be appropriate to discuss briefly the meaning of the 
terms 'Superspace' and space-time interaction. The state 
'Superspace' is the state before the creation of the Universe. 
By the universe, I mean the whole aggregate of objects consisting 
of matter or radiation or both. The state where neither matter 
nor radiation existed is called thc 'Superspace'. The energy of 
this state is the source for the creation of the Universe and 
therefore it is called the fundamental energy. Superspace by 
its definition acquires certain peculiar characteristics. We 
know that identical objects are indistinguishable, the perfect 
symmetry requires the absence of any discontinuity and measurement 
loses its meaning for infinite extent. The concepts of 
dimensionality, origin and boundary are, therefore, meaningless 
for the state 'Superspace'. The fundamental energy is uniformly 
distributed throughout the 'Superspace'. The second postulate 
gives a special place to time as a dynamical operator. The 
fundamental energy associated with the 'Superspace' is transformed 
to some radiation, when time operates upon the 'Superspace'. 
The space time interaction, as I call it, where started, can 
not be asked because of the perfect symmetry of 'Superspace'. 
However, there was certainly a beginning for our Universe, 
because every change has beginning. Moreover, once the interaction 
takes place, a finite, bounded and well defined region will come 
into existence. The measurement sense of both space and time also 
enters into picture as soon as space-time interaction takes place. 
That the energy conversion took place in zero time or instantaneously 
has no meaning implies that some finite elementary time interval, 
$\tau$, exists during which fundamental energy gets transformed into 
some radiation. The Nature should not be irregular and partial 
for any specific time then implies that in each time interval $\tau$, 
an amount of energy, $F$ gets converted to some radiation. 

Thus the elementary time interval, $\tau$, is a basic and absolute 
unit of time provided by the Nature. Further the product of $F$ 
and $\tau$ is also constant. Now the converted energy in the 
form of radiations should be contained somewhere. Let us make the 
simplest and natural assumption to regard 
this spatial region as a sphere of volume $V_e$. Since fundamental 
energy is uniformly distributed and an equal amount of energy, 
$F$ is converted in each time interval, $\tau$, the volume of the Universe 
also increases in the steps of $V_e$. If the energy density is 
denoted by $\rho$, then $\rho V_e = F$. All the time intervals are 
the integral multiples of the elementary time interval but, for 
convenience, we replace the discrete time by a continuous time. 
Then at time $t$, the volume of the Universe is given by 
$V = V_e t/\tau$. The radius, $R$, and the rate of radial expansion, 
$v$, are easily calculated to be 
\begin{eqnarray}
R &=& K t^{1/3}  \nonumber \\
v &=& \frac{dR}{dt} = \frac{1}{3} K t^{-2/3}
\end{eqnarray}
where $K = (3V_e/4\pi\tau)^{1/3}$. Neglecting the second and higher 
order terms in $\tau/t$, the expression for the incremental radius
$\Delta R$ from time $t$ to $t + \tau$ is given by 
\begin{eqnarray}
\Delta R = \frac{1}{3}K \tau t^{-2/3}.
\end{eqnarray} 

Assuming that $F \time \tau = h$, Planck's constant, for the typical 
values of $\rho = 1.9 \times 10^{-9}$ erg cms$^{-3}$, 
$\tau = 5.4 \times 10^{-44}$ sec and age of the Universe, 
$t = 2 \times 10^{10}$ years, we get the 
following results: $V_e \approx 6.5 \times 10^{25}$ cm$^3$, 
$R \approx 5.65 \times 10^{28}$ cm and 
$v \approx 2.997 \times 10^{10}$ cm/sec. 
If we identify $\Delta R$ with 
the Planck's length, $(h G/2\pi c^3)^{1/2}$ where $G$ is gravitational 
constant and $c$ is the velocity of light, then the value 
of $G$ is calculated to be $6.707 \times 10^{-8}\; 
{\rm cm^3 gm^{-1} sec^{-2}}$. 
The radius of the Universe and the gravitational constant $G$ 
calculnted here are in reasonable agreement with their 
values calcurated in other models. If we 
take $\tau = 4.4 x 10^{-24}$ sec, then it turns out that 
$R \approx 10^{14}$ cm, $v \approx 10^{-8}$ cm/sec and 
$G \approx  10^5\; {\rm cm^3 gm^{-1} sec^{-2}}$. 
Therefore it seems that the value 
of $\tau$ equal to $5.4 \times 10^{-44}$ sec is close to the actual 
value of $\tau$ , though the uncertainty in other parameters 
$\rho$ and age of the Universe restricts the validity of 
this value of $\tau$ also. 

Since $\Delta R$ is not constant, we note that the Gravitational 
constant varies with time. If $c$ is assumed to be constant, 
then $G \propto t^{-4/3}$. We must point out here that according to 
Dirac's large number hypothesis \cite{2} $G \propto t^{-1}$. 

The most interesting result is that $v \approx c$. Does it mean 
that the velocity of light is not constant, because 
$v \approx t^{-2/3}$ ? The velocity of light dictates every 
physical process, though it is not clear, why. 
Einstein himself was aware of this criticism for giving 
such a fundamental role to the velocity of light \cite{3}. 
Here we have an explanation for it. The radial expansion rate 
determines the maximum attainable velocity in the Universe. 
It is very I satisfying principle indeed, in our model 
because outside the physical spatial 
boundary of our Universe, there is the 'Superspace' 
containing fundamental energy. Now no energy signal can 
penetrate this ( Superspace' because it is incompressible 
~nd nonempty. Therefore. the restriction on the velocity 
of any signal that it can not exceed the radial expansion 
rate will make it impossible to cross the boundary of 
the Universe. Therefore the physical processes including 
the velocity of light-are guided bya cosmic principle. 

In the simple model, one can easily calculate that the 
velocity of light at $t = 10^{10}$ yrs. was 
$4.7579739 \times 10^{10}$ cm/sec and it will be 
$2.287395 \times 10^{10}$ cm/sec. at $t = 3 \times 10^{10}$ yrs. 

The variation of the light velocity with time will alter 
the concepts of fundamental constants. Dirac \cite{4} argued 
that not all the three fundamental constants $h/2\pi,\; c\; 
{\rm and}\; e$ 
are fundamental. at least one is derived. If the fine 
structure constant is considered independent of time 
then e should also vary with time, since $h/2\pi$ is fundamental 
in our model. Dirac however insisted that $c$ plays such 
an important role in four dimensional picture that it 
must be fundamental. Only the experiments can decide. 
but we expect so many new and interesting consequences 
will follow in case we consider that the velocity of 
light is not constant. 

Finally we would like to give a few remarks regarding 
the origin of the gravitational force. One can see that 
$G \propto t^{-10/3}$, as $v = c \propto t^{-2/3}$. Now, is gravity a real 
field ? Let us discuss it in our model. It is now a well 
established fact that matter energy can be converted to 
radiation energy and vice versa. We now invoke the 
principle of naturality and condensation and conservation 
principle of energy. 

The former demands that every physical system has a 
tendency to attain its original state. When applied to the 
present situation, we find that because the 'Superspace' 
had uniform and homogeneous distribution of fundamental 
energy, the energy in the Universe always has a tendency 
to be uniformly distributed throughout the space. The 
condensation and conservation principle can be stated as 
follows. The energy conversion takes place in such a manner 
that the energy density increases, i.e. fundamental energy 
density $<$ radiation energy density $<$ matter energy density. 
The source of the matter and radiation in the Universe is 
the fundamental energy therefore. local energy conservation 
principle is a necessary corollary of our hypothesis 
because we can not get more energy than what is contained in 
'Superspace'. When radiation and matter are created, the 
empty space is created inside the Universe. The naturality 
principle then brings out a pressure kind of thing which 
manifests as an attractive force. Thus qualitatively we 
can understand the gravitational force, which seems to be 
an apparent interaction. 

Although a number of problems such as red shift, microwave 
back-ground radiation etc. have not been touched upon and 
the model presented here is far from being complete in the 
sense that the details of the large scale structure of 
Univarse are not discussed, the conceptual and philosophical 
profoundness of this model are its attractive features.

\end{document}